\documentclass[journal]{IEEEtran}
\ifCLASSINFOpdf
\else
\fi
\usepackage[dvipsnames]{xcolor}
\usepackage[normalem]{ulem}
\usepackage{color,soul}
\usepackage{enumitem}
\usepackage{comment}
\usepackage{longtable}
\usepackage{graphicx}
\usepackage[font=normalsize]{caption}
\usepackage{chemformula}
\usepackage{caption}
\usepackage{booktabs}
\DeclareCaptionLabelSeparator{tableNewline}{\par}
\usepackage{caption}
\usepackage{subcaption}
\usepackage[font={small},skip=10pt]{caption}
\usepackage{booktabs}
\usepackage[english]{babel}
\usepackage{multicol}
\usepackage{amsmath}
\usepackage{fancyhdr}
\usepackage[section]{placeins}
\usepackage{multirow}
\usepackage{flushend}
\usepackage{tabularx}
\usepackage{adjustbox}
\usepackage{float}
\usepackage{makecell}
\usepackage{booktabs}
\usepackage{array}
\usepackage{hyperref}
\usepackage{tikz}
\usepackage{quantikz}
\usepackage{amsmath, amssymb} 
\hypersetup{
  colorlinks=true,
  allcolors=blue,
  allbordercolors={blue!50!black}}
\usepackage[noadjust]{cite}
\usepackage[english]{babel}
\usepackage{graphicx}
\newcommand{\bluecite}[1]{\textcolor{blue}{\cite{#1}}}

\begin{document}
\bstctlcite{IEEEexample:BSTcontrol}
\setlength{\parskip}{0pt}
\title{Information Theoretic Modeling of Interspecies Molecular Communication}

\author{Bitop Maitra,~\IEEEmembership{Graduate Student Member,~IEEE},
        Murat Kuscu,~\IEEEmembership{Member,~IEEE},
        and~Ozgur~B.~Akan,~\IEEEmembership{Fellow,~IEEE}
\thanks{The authors are with the Center for neXt-generation Communications (CXC), Department of Electrical and Electronics Engineering, Ko\c{c} University, Istanbul 34450, Turkey (e-mail: \{bmaitra23, mkuscu, akan\}@ku.edu.tr).}
\thanks{B. Maitra and M. Kuscu are also with the Nano/Bio/Physical Information and Communications Laboratory (CALICO Lab), Department of Electrical and Electronics Engineering, Ko\c{c} University, Istanbul 34450, Turkey}
\thanks{O. B. Akan is also with the Internet of Everything (IoE) Group, Electrical Engineering Division, Department of Engineering, University of Cambridge, Cambridge CB3 0FA, UK (email: oba21@cam.ac.uk).}
\thanks{This work was supported by the AXA Research Fund (AXA Chair for Internet of Everything at Ko\c{c} University).}

\vspace{-1.2mm}

}

\maketitle

\begin{abstract}
Plants and insects communicate using chemical signals like volatile organic compounds (VOCs). 
A plant encodes information using different blends of VOCs, which propagate through the air to represent different symbolic information.
This communication occurs in a noisy environment, characterized by wind, distance, and complex biological reactions. 
At the receiver, cross-reactive olfactory receptors produce stochastic binding events whose discretized durations form the receiver observation.
In this paper, an information-theoretic framework is developed to model interspecies molecular communication (MC), where receptor responses are modeled probabilistically using a multinomial distribution.
Numerical results show that the communication depends on environmental parameters such as wind speed, distance, and the number of released molecules. 
The proposed framework provides fundamental insights into the VOC-based interspecies communication under realistic biological and environmental conditions.
\end{abstract}


\renewcommand{\figurename}{Fig.}

\section{Introduction}
Communication through chemical signals is a fundamental mechanism by which living systems exchange information, which spans from intracellular signaling to interspecies interactions. 
Considering plant-based communication, plants employ volatile organic compounds (VOCs) as airborne Information Molecules (IMs) as a reaction to biotic/abiotic stresses or as natural constitutive emissions to communicate and convey their status to their nearby plants or species.
In response to the received VOCs, the surrounding organisms are influenced, which includes attracting pollinators or predators, repelling herbivores, etc. \bluecite{Dudareva2006}. 
Unlike electromagnetic communication, VOC-based communication operates in stochastic transport, inconsistent wind flow, and nonlinear biochemical interactions, along with noisy sensing, making it a competitive domain of research.

Molecular Communication (MC) is an emerging domain that studies information transfer using chemical carriers from an Information and Communication Technology (ICT) perspective \bluecite{Nakano2013}. 
Prior work has investigated diffusion-based channels, ligand–receptor binding models, and capacity limits under idealized assumptions such as single-molecule signaling or non-interfering receptors \bluecite{Pierobon2013}. 
However, biological VOC signaling violates these assumptions in several critical ways. 
First, plants emit \emph{blends} of VOCs rather than a single molecule type.
Second, olfactory receptors exhibit \emph{cross-binding}, where multiple VOCs can bind the same receptor and individual VOCs can bind to multiple receptors \bluecite{Firestein2001}.
The information is extracted from the statistics of VOC-receptor binding durations rather than the instantaneous molecular count.
Third, airborne propagation occurs in an ambient environment dominated by advection–diffusion dynamics rather than being diffusion-limited \bluecite{Okubo2001}.

\begin{figure}
    \centering
    \includegraphics[width=0.9\linewidth]{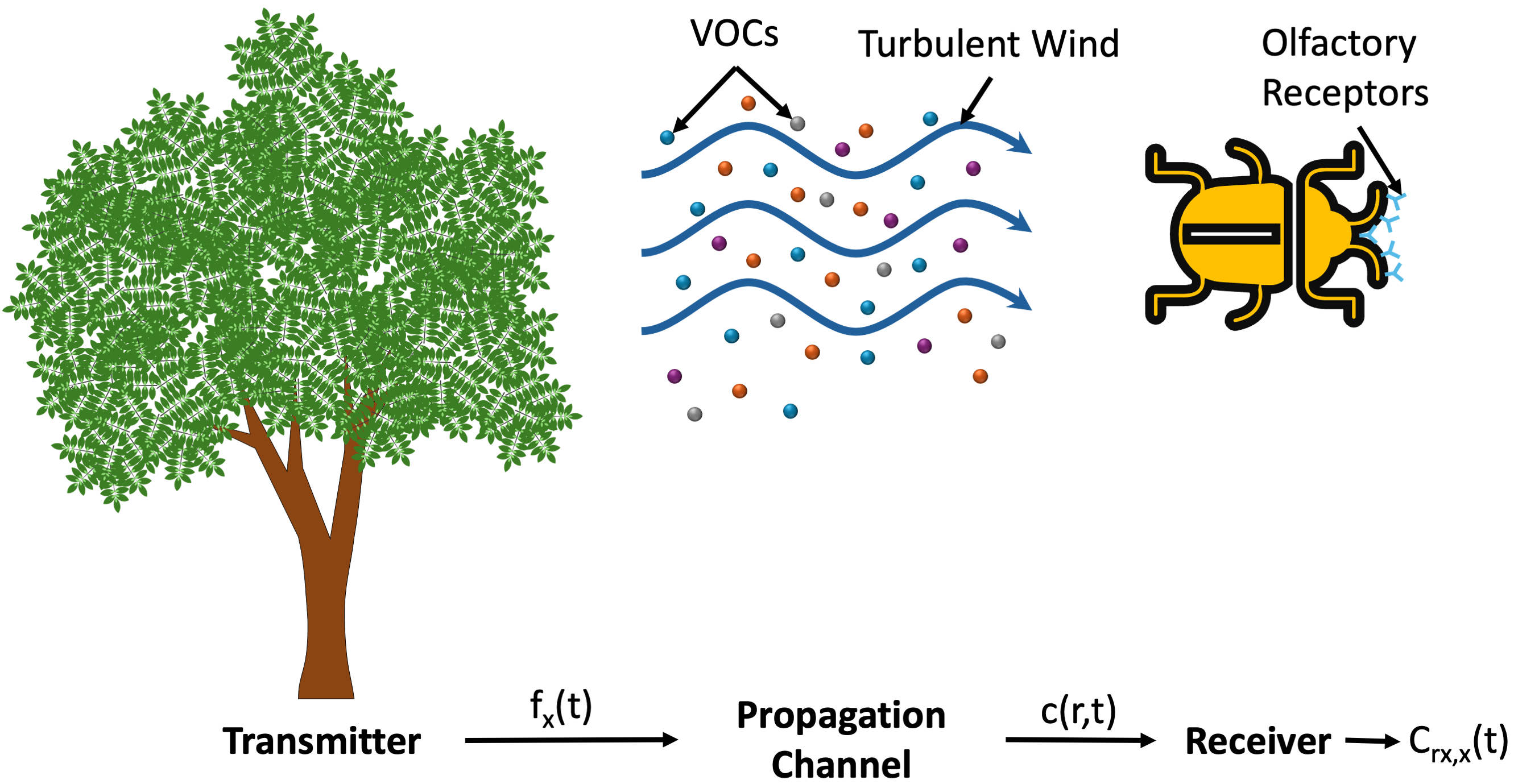}
    \caption{Schematic of Interspecies Molecular Communication.}
    \vspace{-2.5mm}
    \label{fig:placeholder}
\end{figure}
In this work, an ICT framework for interspecies MC is developed based on symbolic VOC emission. 
A plant transmitter is considered that encodes binary information through distinct VOC blends, where symbol~$0$ represents attraction of pollinators and symbol~$1$ represents repulsion of herbivores. 
The emitted VOC molecules propagate through the environment according to an advection-diffusion process, capturing the combined effects of wind-driven transport and molecular dispersion. 
As a result, the VOC concentrations at the receiver evolve continuously over time within each symbol interval.

The receptor layer is modeled as a cross-reactive molecular channel in which the received VOC concentration governs the stochastic binding events.
Information is extracted from the durations of these binding events, which are discretized into time bins to form the receiver observation.
Conditioning on the total number of binding events within a symbol interval, the resulting binding-time bin count vector follows a multinomial distribution determined by VOC-receptor kinetics.
This probabilistic mapping defines a discrete-time memoryless channel whose transition statistics depend on the propagation distance, VOC blend composition, and receptor kinetics.
Using an information-theoretic formulation based on Fisher Information (FI), the asymptotic capacity is characterized for VOC-based MC under realistic environmental and biochemical conditions. 

The main contributions of this paper are as follows. 
First, a binding time-based receiver model for interspecies communication is modeled, where information is extracted from receptor binding durations. 
Second, cross-reactive receptor observations using a multinomial binding-time bin framework are explored. 
Finally, we develop an information-theoretic analysis based on FI to characterize the impact of propagation conditions, including wind speed and distance. 

The rest of the paper is organized as follows.
Sec. \ref{sec:model} presents the system model, including symbolic VOC emission, advection–diffusion propagation, cross-receptor binding dynamics at the receiver, and the time-averaged asymptotic capacity based on FI. 
Sec. \ref{sec:results} presents numerical analysis, which discusses the impact of transmission strength, distance, and flow velocity on interspecies MC.
Finally, Sec. \ref{sec:conc} concludes the paper with the future direction.

\vspace{-4mm}
\section{System Model}
\label{sec:model}
\subsection{Transmission Model}
Plants use multiple VOCs to convey information to nearby plants \bluecite{maitra2025modeling} and surrounding insects \bluecite{proffit2020chemical} or animals. 
The receiver interprets the information based on the blend of the VOC mixture and its concentration level.
In this study, we consider the transmission of information symbolically, where attracting pollinators and repelling herbivores are considered to be two symbols, where a blend of different VOCs is released according to the symbol.
The released VOC type widely varies according to the plant species, which are heavily modulated by the biotic or abiotic stresses, and environmental requirements.
The resulting airborne signals are interpreted in a receiver-specific manner, which is influenced by the chemical and binding properties of the receptors.

In this study, we consider a total of 6 VOCs, which create the symbols of our interest.
VOCs such as $\beta$-Ionone \bluecite{paparella2021beta}, MeSA, and Limonene \bluecite{das2013plant} can function as both a pollinator attractant and a herbivore repellent depending on the plant species and stress type.
Moreover, Geraniol dominantly acts as a pollinator attracting \bluecite{slavkovic2023floral}, whereas, $\beta$-Caryophyllene and $\beta$-Ocimene primarily act as repelling herbivores \bluecite{heil2014herbivore, war2012mechanisms}.

The transmitter is modeled as a burst release of VOCs.
Two symbols ($x \in \{0,1\}$), which are represented by the different blends of VOCs, have an emission rate of $f_x(t)$ over a symbol interval of [0, $T_s$].
The information is encoded using VOC blend modulation. 
Each symbol is represented by a fixed VOC fraction vector, given by $\rho_x = [\rho_{x,1},\dots, \rho_{x,M}]$, where M is the number of VOCs and $\sum_{i=1}^M \rho_{x,i} = 1$.
During each $T_s$, a fixed total number of molecules $N$ is released, and the number of molecules of VOC $i$ is given by $N_{x,i} = N\rho_{x,i}$,
and propagated through the medium.
To characterize the channel generally, a continuous blend parameter $s\in[0,1]$ is introduced, where $s=0$ and $s=1$ correspond to the two extreme VOC blends $\rho_0$ and $\rho_1$, respectively.

\subsection{VOC Propagation Channel Model}
After the release of VOCs from the transmitter plant, the VOCs travel through the ambient air and propagate towards the receiver, i.e., an insect. 
The propagation of VOCs, considering the advection–diffusion environment in the ambient air, is modeled.

Some assumptions are imposed for the modeling of the channel. 
Firstly, the velocity of the wind ($u_w$) and the velocity of the insect ($v_r$) are constant in the medium. 
Moreover, the movement of the insect is restricted to 1 dimension only in this study.
Lastly, the insect's velocity should be less than the speed of the wind.

The emitted VOCs from the transmitter start to propagate in the medium as given by the diffusion-advection equation \bluecite{kuscu2018modeling}: 
\begin{align}
    \frac{\partial c (\Vec{r}, t)}{\partial t} = D \nabla^2 c(\Vec{r},t ) - u \cdot \nabla  c(\Vec{r},t) + S (\Vec{r},t),
    \label{diff-adv}
\end{align}
where $c(\Vec{r},t)$ is the concentration at a position in 3D, $D$ is the diffusion constants of VOCs, $u$ is the relative velocity, expressed as $u = u_w-v_r$, $\nabla$ and $\nabla^2$ are the gradient and Laplacian operator, respectively, and $S (\Vec{r},t)$ is the source term.

The impulse response of the (\ref{diff-adv}) is widely studied in the literature for MC \bluecite{maitra2024molecular}, and expressed as:
\begin{align}
    c_{rx}(t) = \frac{N}{{(4\pi Dt)}^{3/2}} exp\left(-\frac{(d_0-u t)^2}{4Dt}\right),
    \label{eq:impulse}
\end{align}
where $d_0$ is the position of the insect at time $t_0$, $t$ is the elapsed time since emission.

\begin{figure}
    \centering
    \includegraphics[width=\linewidth]{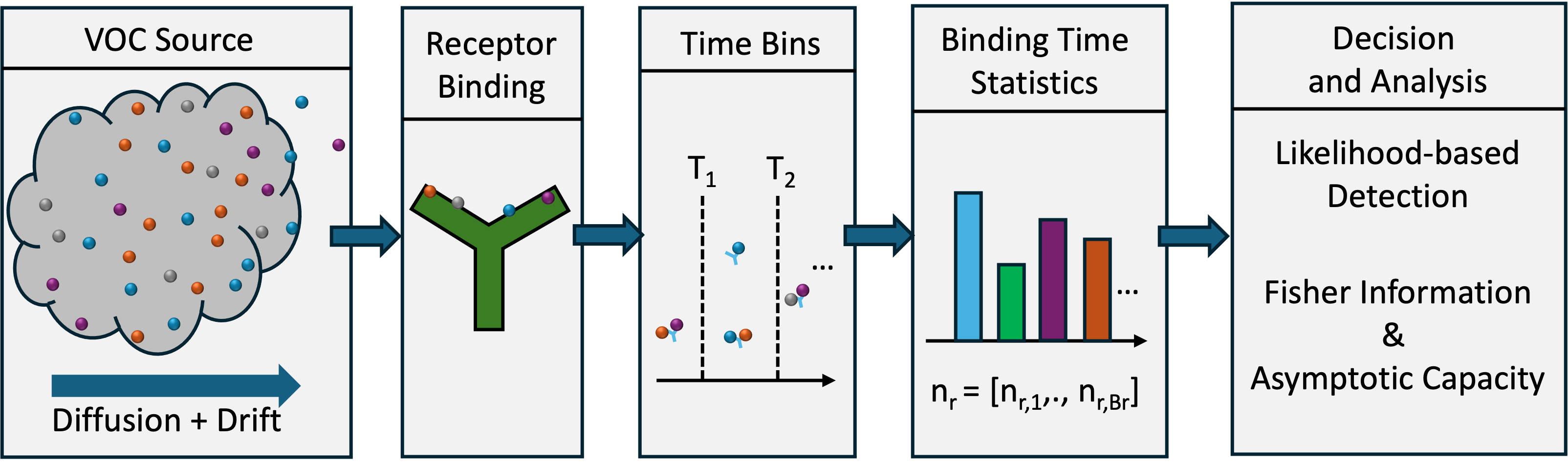}
    \caption{End-to-end receiver observation and decision pipeline based on multinomial binding-time bin count statistics.}
    \label{fig:end-end}
\end{figure}

\subsection{Insects Reception Model}
Considering the release of VOCs and the channel impulse response, the concentration at the receiver can be expressed as the convolution of them. The received concentration is obtained by the linear superposition of time-shifted per-symbol emission contributions, expressed as:
\begin{align}
   C_{rx,x} (t) = \sum_k\int_{t_k}^{t_k+T_s} &f_x(\tau-t_k) \frac{1}{({4 \pi D (t-\tau)})^{3/2} } \nonumber \\  & exp (-\frac{(d_0-u(t-\tau))^2}{4D(t-\tau)}) d\tau ,
   \label{rec_conc}
\end{align}

\begin{table*}[]
    \centering
    \caption{Diffusion coefficient ($D (cm^2/s)$) and binding rate ($k^+ (M^{-1}s^{-1})$) of different VOCs}
    \label{tab:diffu}
    \begin{tabular}{c c c c c c c}
    \hline
        VOC & $\beta$-Ionone & MeSA & Geraniol & Limonene  & $\beta$- Caryophyllene & $\beta$-Ocimene \\ \hline
        $D$ & 0.0433 \bluecite{burgess2024self} & 0.0816 \bluecite{tang2015compilation} & 0.0432 \bluecite{burgess2024self} &	0.0584 \bluecite{burgess2024self}	& 0.0439 \bluecite{burgess2024self} &	0.0584 \bluecite{burgess2024self} \\
        $k^+$ & 5.21E+13 &	9.82E+13 &	5.20E+13 &	7.04E+13 &	5.29E+13 &	7.04E+13 \\ \hline
    \end{tabular}
\end{table*}

\begin{table*}[t!]
\scriptsize
\centering
\caption{Dissociation constants ($K_D$ ($\mu$M)) and unbinding rate ($k^{-}$ (s$^{-1}$)) between VOC and receptor}
\label{tab:binding-constants}
{\fontsize{6}{8}\selectfont
\begin{tabular}{l *{6}{c c}}
\toprule
\multirow{2}{*}{Receptor} & 
\multicolumn{2}{c}{$\beta$-Ionone} & 
\multicolumn{2}{c}{MeSA} & 
\multicolumn{2}{c}{Geraniol} & 
\multicolumn{2}{c}{Limonene} & 
\multicolumn{2}{c}{$\beta$-caryophyllene} & 
\multicolumn{2}{c}{$\beta$-ocimene} \\
\cmidrule(lr){2-3} \cmidrule(lr){4-5} \cmidrule(lr){6-7} \cmidrule(lr){8-9} \cmidrule(lr){10-11} \cmidrule(lr){12-13}
& $K_D$ & $k^{-}$ & $K_D$ & $k^{-}$ & $K_D$ & $k^{-}$ & $K_D$ & $k^{-}$ & $K_D$ & $k^{-}$ & $K_D$ & $k^{-}$ \\
\midrule
ApisOBP9 & 5.87 \bluecite{pan2024sustainable} & 3.06E+8 & 7.74 \bluecite{yang2023aphid} & 7.60E+08 & 9.10 \bluecite{yang2023aphid} & 4.73E+08 & $-$ & $-$ & 12.50 \bluecite{pan2024rational} & 6.62E+08 & $-$ & $-$ \\
HaxyOBP15 & 5.15 \bluecite{pan2024sustainable} & 2.69E+08 & 13.35 \bluecite{yang2023aphid} & 1.31E+09 & 12.69 \bluecite{yang2023aphid} & 6.60E+08 & $-$ & $-$ & $-$ & $-$ & $-$ & $-$ \\
ApisOBP2 & $-$ & $-$ & $-$ & $-$ & $-$ & $-$ & 56.10 \bluecite{wang2019molecular} & 3.95E+09 & 0.79562  \bluecite{wang2019molecular} & 4.21E+07 & 150.95 \bluecite{wang2019molecular} & 1.06E+10 \\
\bottomrule
\end{tabular}
}
\end{table*}

In this study, the receiver is modeled as extracting statistical information of VOC-receptor interaction, without assuming a deterministic symbol-level decoding mechanism.
A schematic of the end-to-end receiver observation and decision pipeline is depicted in Fig. \ref{fig:end-end}.
Due to cross-reactive receptors, information can not be reliably encoded or decoded using concentration or molecule counts.
Instead, insect olfactory systems use VOC-specific receptor binding durations as the primary information carrying variable, and the receiver extracts information from the binding duration statistics. 
Considering the binding mechanism, VOCs bind with receptors. 
We are dividing the molecular binding to receptors into non-overlapping intervals, separated by thresholds according to the VOCs, given by $T_i = \frac{\nu}{k_i^-}$, where $k_i^-$ denotes the unbinding rate of VOCs \bluecite{kuscu2022detection}. 
In this paper, we set $\nu=3$, which is shown in \bluecite{kuscu2019channel} to result in near-optimal performance for channel sensing accuracy and estimation error.
These time bins discretize continuous binding durations into a kinetically distinguishable observation space.
If the bound duration of each VOC is given by $\tau_{B,i}$,
and conditioned on the received concentration in \eqref{rec_conc}, the probability that a binding event falls into a given binding-time bin is given by \bluecite{kuscu2022detection}:
\begin{align}
    p(\tau_{B,i}) = \sum_{i \in VOC} \rho_{x,i} k_i^{-}e^{-k_i^-\tau_{B,i}},
    \label{binding}
\end{align}
Hence, over $T_s$, each receptor observes a histogram of binding events across time bins for different VOC blends.
The probability of binding within a binding duration can be expressed in matrix form as $p_r = Q_r \rho_s$, and $Q_r$ is given by:

{\footnotesize
\[
Q_{r} = \\ \begin{pmatrix}
1-e^{-k_1^{-}T_1} &  ... & 1-e^{-k_n^{-}T_1} \\
e^{-k_1^{-}T_1}-e^{-k_1^{-}T_2}  & ... & e^{-k_n^{-}T_1}-e^{-k_n^{-}T_2}  \\
... & ... & ...  \\
... & ... & ...  \\
e^{-k_1^{-}T_{m-1}}-e^{-k_1^{-}T_{m}} & ... & e^{-k_n^{-}T_{m-1}}-e^{-k_n^{-}T_{m}}  \\
e^{-k_1^{-}T_m}  & ... & e^{-k_n^{-}T_m} 
\end{pmatrix}_{m \times n}
\]
}
\noindent where $m$ is the number of time bins, and $n$ is the number of VOCs attach to receptors.

The observed bound receptors follow a multinomial distribution, expressed as:
\begin{align}
    n_r \mid x \sim \mathrm{Multinomial}\big(N_r(x), p_r^{(x)}\big),
\end{align}
where $N_r(x)$ is the number of binding events encountered by receptor $r$ during one symbol interval, and $n_r$ is the number of $N_r(x)$ bindings that fall into each bin of receptor.
Thus, the effective binding observation of receptor $r$ during $T_s$ is $n_r = [n_{r,1},\dots, n_{r,B_r}]$, where each entry is the number of binding events whose duration falls within the corresponding time bin $\in {1, \dots, B_r}$.
Discretizing the binding time distribution in \eqref{binding} and using the time bins, the bin probability $p_{r,b}^{(x)}$ is obtained. 
Hence, multinomial PMF is expressed as:
\begin{align}
    P(n_r \mid x)=\frac{N_r(x)!}{\prod_{b=1}^{B_r} n_{r,b}!} \prod_{b=1}^{B_r} \left[p^{(x)}_{r,b}\right]^{n_{r,b}},
    \label{multinomial}
\end{align}
where $n_{r,b}$ is the number of dissociation events from a receptor $r$ time bin $b$, $p_{r,b}^{(x)}$ is the probability that a binding to a receptor $r$ within time bin $b$.
It is worth noting that the model is focused on receptor-level observations and does not consider the subsequent neural processing.

Although information is transmitted using discrete symbols, the multinomial likelihood is parametrized by $s$ to evaluate Fisher Information \bluecite{komorowski2025closed}, which characterizes the local sensitivity of the likelihood function with respect to the VOC blend parameter.
Considering the log-likelihood, (\ref{multinomial}) reduces to:
\begin{align}
    \ell_r(n_r \mid s)=\log \big(N_r(x)!\big) &- \sum_b\log \big(n_{r,b}!\big) \nonumber \\
    &+ \sum_b n_{r,b} \log p_{r,b}(s),
    \label{log}
\end{align}
where $p_{r,b}{(s)} = (1-s) p_{r,b}^{(0)} + s p_{r,b}^{(1)}$.
Taking the derivative of (\ref{log}) with respect to $s$, we obtain:
\begin{align}
    \frac{\partial \ell_r}{\partial s} = \sum_b \frac{n_{r,b}}{p_{r,b}(s)} \frac{\partial p_{r,b} (s)}{\partial s},
\end{align}
The Fisher Information (FI) matrix is given by \bluecite{komorowski2025closed}:
\begin{align}
    FI_{ij} (s) = \mathbb{E} \left[\frac{\partial \ell_r}{\partial s_i} \cdot \frac{\partial \ell_r}{\partial s_j}\right],
    \label{F_ij}
\end{align}
where $i$, $j$ are the independent blend-composition degrees of freedom.
Considering the multinomial receptor binding with probability $p_{r,b}(s)$, the (\ref{F_ij}) reduces to:
\begin{align}
    FI_r(s) = N_r (x) \sum_b \frac{1}{p_{r,b}(s)} \left(\frac{\partial p_{r,b}(s)}{\partial s}\right)^2,
\end{align}
For multinomial evaluation, a linear interpolation path in probability space is assumed, s.t., $\frac{\partial p_{r,b}(s)}{\partial s} = \Delta p_{r,b}$.
As all the receptors are working independently, considering i.i.d. receptors, the total FI becomes:
\begin{align}
    FI_{total}(s) = N_r \sum_r \sum_b \frac{(\Delta p_{r,b})^2}{p_{r,b}(s)},
\end{align}
where $\Delta p_{r,b} = p_{r,b} ^{(1)} - p_{r,b} ^{(0)}$.
The FI based sensitivity metric is expressed as \bluecite{komorowski2025closed}:
\begin{align}
    V & = \sqrt{FI_{total} (s)},
\end{align}

Although information is transmitted using discrete symbols, the stochastic receptor induces a continuous probability distribution via VOC blending; hence, we employ capacity formulation based on FI.
The asymptotic capacity is expressed as \bluecite{komorowski2025closed}:
\begin{align}
    C_A & = log_2 \left[(2 \pi e)^{-l/2} \int_0^1 V ds \right] ,
    \label{C_A}
\end{align}
where $l$ is the dimensionality of the probability simplex formed by the receptor binding events, defined as $l=\sum_rB_r-1$ \bluecite{amari2000methods}, where $\sum_rB_r$ is the total number of time bins for all receptors.

Following this, to express the capacity as a rate, normalizing \eqref{C_A} by the symbol duration $T_s$, the time-averaged asymptotic capacity can be expressed as:
\begin{align}
    C_{A_v} = \frac{1}{T_s}log_2 \left[(2 \pi e)^{-6} \int_0^1 \sqrt{FI_{total} (s)} ds  \right],
\end{align}

\section{Numerical Results and Discussion}
\label{sec:results}
In Sec. \ref{sec:model}, the transmission from the plant, the propagation channel, and the insect's reception processes are described and modeled.
In this section, a numerical analysis of the described system model is done, and the model performance is analysed from the ICT perspective.

\begin{figure}[t!]
    \centering
    \includegraphics[width=0.9\linewidth]{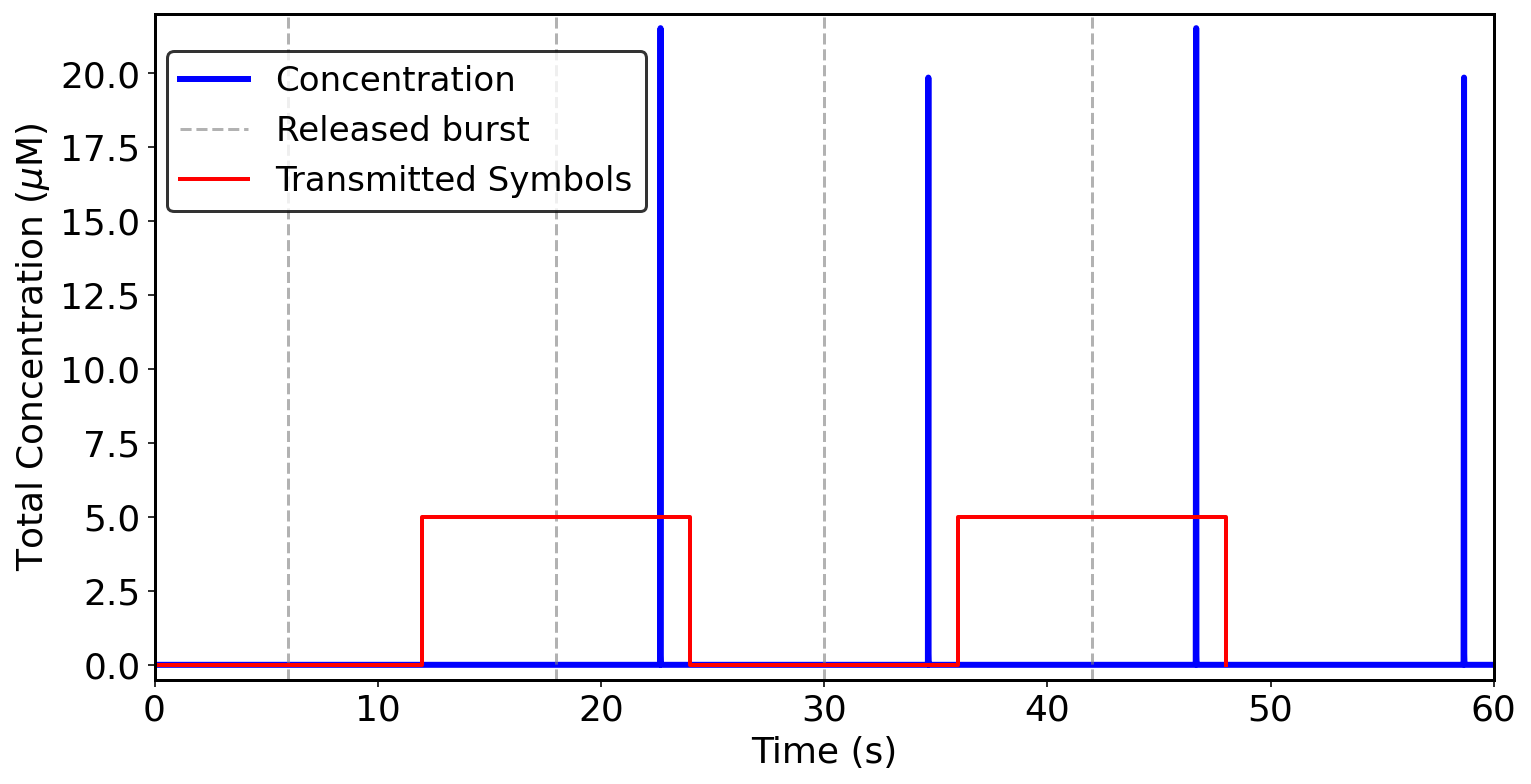}
    \caption{Temporal profile of total VOC concentration with $N$=1 $\mu$mol, $u_w$=700 cm/s and $v_r$=200 cm/s at the receiver (blue) for the corresponding transmitted binary symbol sequence (red square wave) along with emission times of VOC bursts (grey dashed lines).}
    \label{fig:receiver_conc}
    \vspace{-2mm}
\end{figure}
As mentioned earlier, we have considered 6 VOCs in this study, namely $\beta$-Ionone, Methyl-salicylate (MeSA), Geraniol, Limonene, $\beta$-Caryophyllene, and $\beta$-Ocimene.
Diffusion coefficient ($D$) and binding rate ($k^+$) of the VOCs are tabulated in Table \ref{tab:diffu}.
However, it is noteworthy that $D$ and $k^+$ are not widely studied in the literature. 
$D$ is obtained by considering chemical class type (e.g., alkane, alkene, benzene, etc.), and the molecular weight, as discussed in \bluecite{burgess2024self}.
The binding rate ($k^+$) is determined via $k^+ = 4Da$, where $a$ is the effective receptor size, considered to be 5 nm.
This is a tolerable consideration for the diffusion-limited MC \bluecite{kuscu2022detection}.
Furthermore, the unbinding rate ($k^-$) is also an important parameter, tabulated in Table \ref{tab:binding-constants}, that influences the capacity.
$k^-$ is obtained in an indirect way by considering the formula $K_D = \frac{k^-}{k^+}$, where $K_D$ is the VOC and receptor specific dissociation constant.
In this study, 3 receptors in an insect are considered: ApisOB9, HaxyOBP15, and ApisOB2.
Moreover, the release of VOCs is modeled according to the symbol, where the common VOCs have an equal percentage of 20\% each, and the remaining percentage is equally divided into the symbol-specific VOC.
Precisely, $s=0$ is represented by $\rho_0 = [0.2,0.2,0.4,0.2,0,0]$, and $s=1$ is considered as $\rho_1 = [0.2,0.2,0,0.2,0.2,0.2]$.

The VOC concentration at the receiver is modeled in (\ref{eq:impulse}), which is depicted via the blue line in Fig. \ref{fig:receiver_conc} with $N$=1 $\mu$mol, i.e., 6.022$\times$10$^\text{17}$ per burst, wind velocity of 700 cm/s, insect velocity of 200 cm/s, and VOC-specific diffusion coefficients $D_i$. 
The symbol-specific VOCs are released burst-wise at $t_k\in\{6,18,30,42\}$ s, which is shown by the gray dashed line, and the red square wave encodes the binary message, with symbol windows $w_i=[12i,12(i+1))$ s, where $i\in{1,2,3,4}$. 
The received concentration and timing at the receiver greatly depend on the speed of the wind, the insect's flying velocity, and the distance between the source and receptor.

\begin{figure}[t!]
    \centering
    \includegraphics[width=0.9\linewidth]{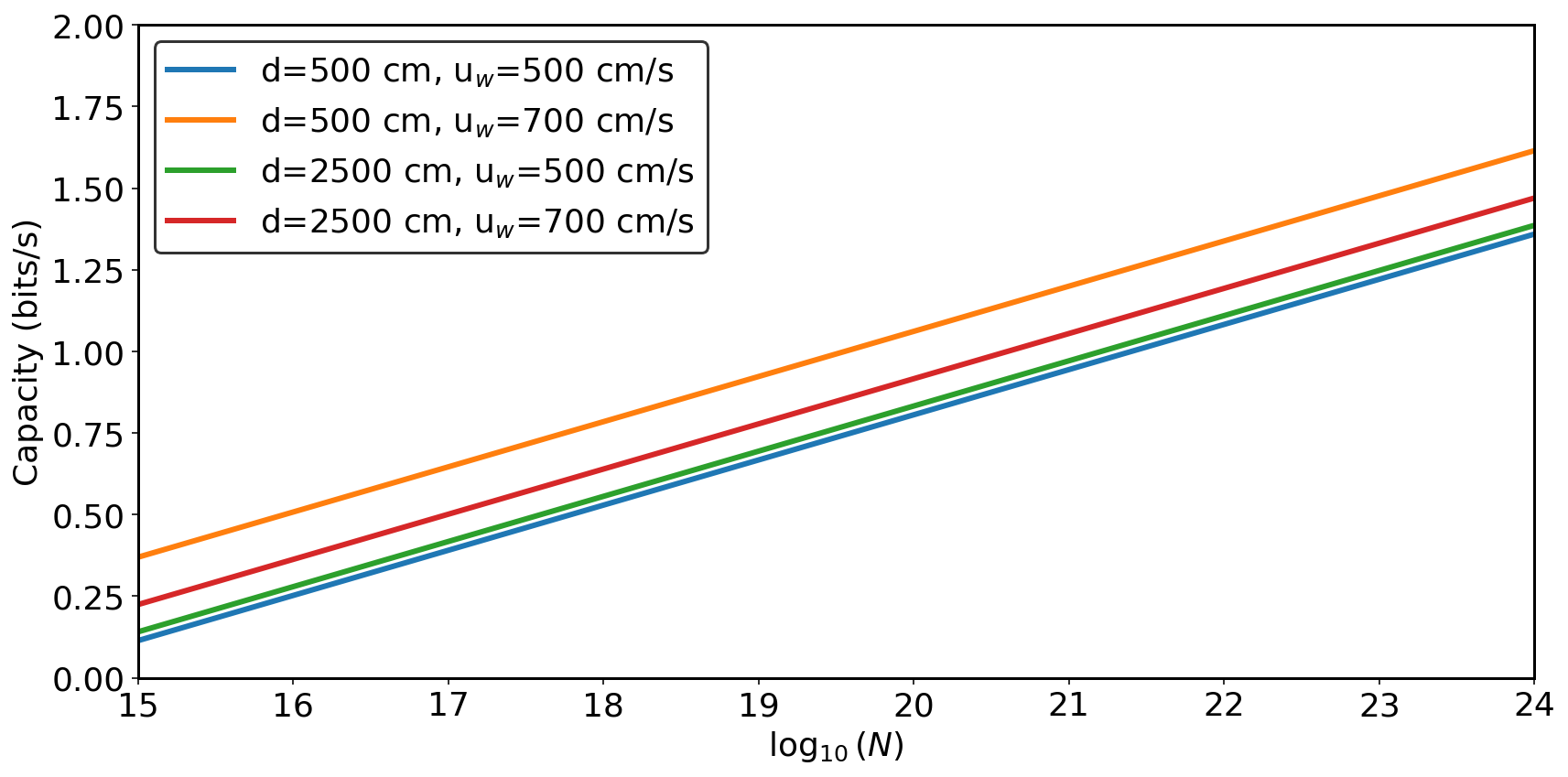}
    \caption{Time averaged asymptotic capacity as a function of number of VOC molecules released per burst $log_{10}(N)$ with $v_r$=200 cm/s.}
    \label{fig:capa_no}
    \vspace{-2.5mm}
\end{figure}
Fig. \ref{fig:capa_no} demonstrates the joint impact of the propagation distance and the flow velocity with respect to the released molecules on MC performance. 
For a fixed distance, higher wind velocity results in higher capacity, as faster drift results in a higher delivery rate of VOCs to the receiver.
Conversely, for a fixed flow velocity, increasing the distance doesn't always reduce capacity.
Even though larger distances introduce greater propagation loss, it broadens the temporal distribution of arriving molecules, resulting in increased receptor-VOCs interaction time, which can lead to higher time-averaged capacity at the moderate distances compared to the very short distances.

\begin{figure}[t!]
    \centering
    \includegraphics[width=0.9\linewidth]{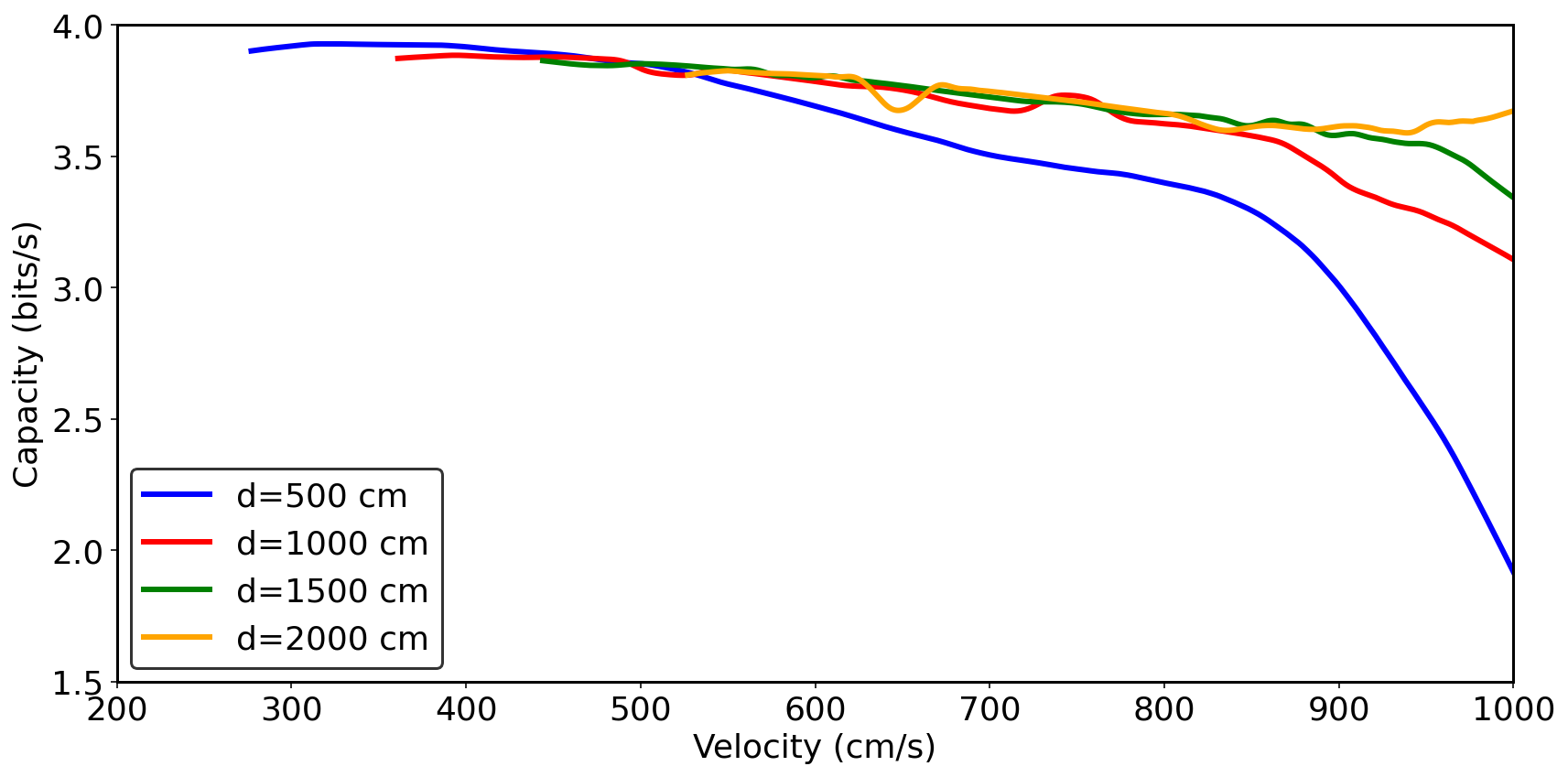}
    \caption{Time averaged asymptotic capacity w.r.t wind velocity for different source–receiver distances with $N$=1$\mu$mol and $v_r$=200 cm/s.}
    \label{fig:capa_u}
    \vspace{-2.5mm}
\end{figure}
Fig. \ref{fig:capa_u} shows how interspecies communication is influenced as the wind velocity increases for several fixed distances. 
At lower speeds, capacity is generally higher, especially for the shorter distances.
When the wind velocity is low, VOCs spend more time near the receptors, increasing the likelihood of binding and allowing receptors to bind VOCs and collect more reliable information during each symbol interval, leading to higher capacity.
As the velocity increases, the capacity gradually decreases or flattens out. 
This is because molecules are transported past the receptors more quickly, reducing the interaction time, even though they arrive faster.
For larger distances, the capacity becomes less sensitive to changes in velocity.
Once the flow is fast enough for the VOCs to reach the receiver, further increase in speed does not improve communication because the signal strength is limited by distance.
Overall, the plot shows that faster transport does not imply better communication; rather, slower wind flow with sustained exposure may be more informative.

\begin{figure}[t!]
    \centering
    \includegraphics[width=0.9\linewidth]{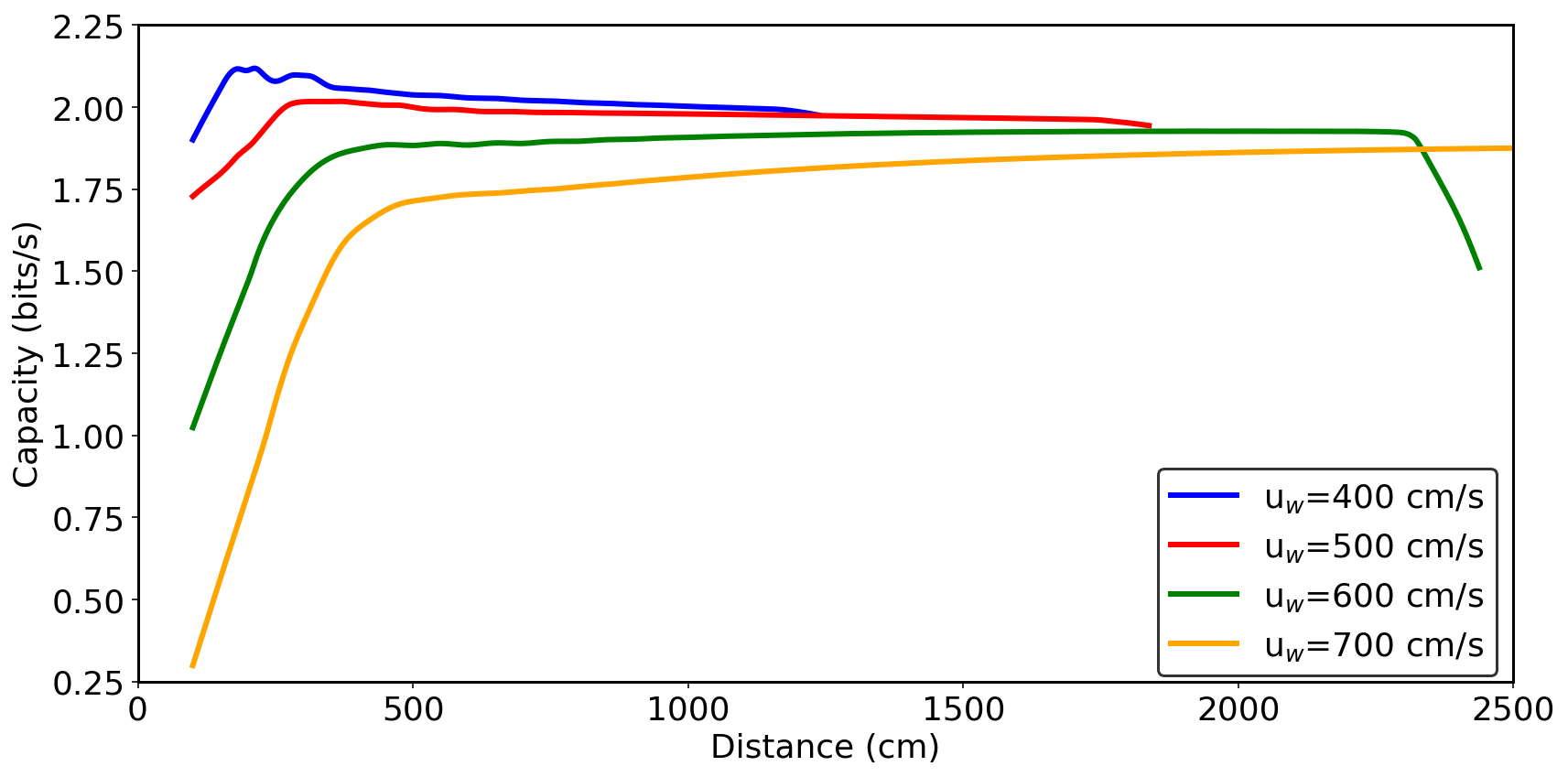}
    \caption{Time averaged asymptotic capacity w.r.t distance for different wind velocity with $N$=1$\mu$mol and $v_r$=200 cm/s.}
    \label{fig:capa_d}
    \vspace{-2.5mm}
\end{figure}
Fig. \ref{fig:capa_d} depicts how the asymptotic capacity varies with respect to distance for different wind velocities. 
When the transmitter is extremely close to the insect, the capacity is relatively low and then increases as the distance increases.
This is because when the transmitter is very close, VOCs arrive quickly in a narrow time window, which limits the number of VOCs that can bind with receptors.
With a small increase in the distance, VOCs get a slightly longer time span, which allows the receptors to bind more effectively, resulting in increased capacity.
As the distance increases, the capacity slowly decreases for lower wind velocities due to the gradual decay in the number of molecules for the VOC spreading process, whereas the capacity increases gradually for higher wind velocities due to the drift induced by faster transport, supporting the long-distance interspecies communication.

\section{Conclusion}
\label{sec:conc}
This paper presents an ICT model for interspecies MC using plant-emitted VOCs. 
Considering advection-diffusion propagation of VOCs and binding time-based cross-reactive receptor observations, this study captures realistic biological and environmental effects.
The analysis shows that communication is not totally determined by faster transport or shorter distances. 
Instead, factors such as receptor interaction time with VOCs, wind velocity, and propagation distance all together influence the achievable time-averaged asymptotic capacity. 
Interestingly, moderate distances and slower wind speeds can sometimes improve communication by increasing the duration of VOC–receptor interactions.
This work bridges biology and communication theory, which lays the groundwork for future studies on multi-symbol signaling, adaptive VOC emission, and more complex interspecies communication networks.

\bibliography{references}
\bibliographystyle{IEEEtran}

\end{document}